# Thermal Oxidation of WSe$_2$ Nano-sheets Adhered on SiO$_2$/Si Substrates

Yingnan Liu,[a,‡] Cheng Tan,[b,c,‡] Harry Chou,[c] Avinash Nayak,[b] Di Wu,[a] Rudresh Ghosh,[c] Hsiao-Yu Chang,[b] Yufeng Hao,[c] Xiaohan Wang,[c] Joon-Seok Kim,[b] Richard Piner,[c] Rodney S. Ruoff,[c,d,e] Deji Akinwande,[b,*] Keji Lai[a,*]

a)  Department of Physics, University of Texas at Austin, Austin, TX, 78712, USA
b)  Microelectronics Research Center, University of Texas at Austin, Austin, TX 78758, USA
c)  Department of Mechanical Engineering and the Materials Science and Engineering Program, University of Texas at Austin, Austin, TX 78712, USA
d)  Center for Multidimensional Carbon Materials, Institute for Basic Science (IBS), Ulsan 689-798, Republic of Korea
e)  Department of Chemistry, Ulsan National Institute of Science and Technology (UNIST), Ulsan 689-798, Republic of Korea

* Correspondence and requests for materials should be addressed to K.L. (kejilai@physics.utexas.edu) or D.A. (deji@ece.utexas.edu)

‡ These authors contributed equally to this work.

## Abstract

**Due to the drastically different intralayer versus interlayer bonding strengths, the mechanical, thermal, and electrical properties of two-dimensional (2D) materials are highly anisotropic between the in-plane and out-of-plane directions. The structural anisotropy may also play a role in chemical reactions, such as oxidation, reduction, and etching. Here, the composition, structure, and electrical properties of mechanically exfoliated WSe$_2$ nano-sheets on SiO$_2$/Si substrates were studied as a function of the extent of thermal oxidation. A major component of the oxidation, as indicated from optical and Raman data, starts from the nano-sheet edges and propagates laterally towards the center. Partial oxidation also occurs in certain areas at the surface of the flakes, which are shown to be highly conductive by microwave impedance microscopy. Using secondary ion mass spectroscopy, we also observed extensive oxidation at the WSe$_2$–SiO$_2$ interface. The combination of multiple microcopy methods can thus provide vital information on the spatial evolution of chemical reactions on 2D materials and the nanoscale electrical properties of the reaction products.**

Layered van der Waals (vdW) materials, in which the electronic properties are inherently anisotropic, have been studied for a relatively long time [1], and have also attracted renewed interest in recent years [1, 2]. This family of materials include elemental atomic sheets (e.g., graphene [3], silicene [4, 5], germanene [5], and phosphorene [6, 7]) and transition metal dichalcogenides [8] (TMDCs, e.g., $MoS_2$ and $WSe_2$), among others, presenting a wide array of candidates for research and applications. Given the extensive knowledge obtained on the solid-state chemistry of conventional Group IV elemental semiconductors and III-V compounds, and the potential use of vdW materials in nanoelectronics, it is anticipated that much effort will be devoted to investigate the spatial and temporal evolution of various chemical reactions [9], such as solution or vapor-based synthesis, reduction and oxidation, wet and dry etching, in 2D materials. Specifically, the understanding of these processes from atomic to mesoscopic length scales will provide important insight on their performance at the device level.

The oxidation process of TMDCs is of particular interest due to its strong influence on the characteristics of devices that are not hermetically sealed and could potentially be oxidized in ambient environments. In addition, the fully oxidized products, e.g., $WO_3$, are semiconducting metal oxides with a band gap in the range of 2.5 – 3.7 eV, which may find applications in many areas [10–12]. To date, the oxidation of TMDCs has been studied at elevated temperatures [13], under intense laser illumination [14], and upon exposure to oxygen plasma [15] or ozone [16]. In contrast to conventional semiconductors such as silicon, the TMDCs show larger oxidative reactivity at the edges and surface defective sites [16, 17]. A careful experimentally analysis combining compositional, structural, and electrical imaging tools, on the other hand, has not been carried out to demonstrate how the oxidation process evolves in space and time.

The $WSe_2$ nano-sheets in this study, with typical thicknesses of 2~100 nm and lateral dimensions of 2~100 μm, were mechanically exfoliated onto 285-nm thick $SiO_2$ on heavily doped Si substrates. The samples were placed in a quartz tube furnace, heated up to $T$ = 400 ºC with a ramp time of 6 minutes, and annealed at 400 ºC for a holding time of $t_h$ under ambient atmosphere. The temperature was chosen such that $SeO_2$, which sublimes above 350 °C, does not stay in the solid phase and complicate the elemental analysis shown below. The relative humidity of the environment was 40~50% at the room temperature. We did not specifically investigate the influence of water vapor in this experiment. The heater was then turned off and the furnace was purged with $H_2$/Ar (2 sccm/100 sccm) for the sample to cool down in a non-oxidizing environment. A detailed *T-t* profile can be found in Supporting Information S1. Fig. 1(a) shows a sequence of snapshots captured by a camera during a continuous heating process over 2 hours. The thermal oxidation, as indicated by the clear optical contrast between $WSe_2$ and $WO_3$, started from the sample perimeter and propagated towards the center. Fig. 1(b) shows the lateral distance from the sample edge to the $WO_3$–$WSe_2$ interface as a function of time in this oxidation process. Compared with the classical Deal-Grove model [18], which is used to explain the oxidation of conventional 3D semiconductors such as silicon, the linear time dependence in Fig. 1(b) indicates that the interface reaction rate is the limiting factor throughout the entire process. More snapshots and a brief analysis of the reaction dynamics can be found in Supporting Information S2. We also acquired cross-sectional transmission electron microscopy (TEM) images of a partially oxidized flake. As seen from the TEM data and their fast-Fourier transformed (FFT) images in Fig. 1(c), the vdW layered structure on the $WSe_2$ side remained largely intact, while 3D polycrystalline lattices appeared on the $WO_3$ side [19]. The TEM data also suggest that the interface is relatively broad, with signatures of both $WSe_2$ and $WO_3$ observed across a width of 10~100 nm.

The structural and electrical properties of thermally oxidized $WSe_2$ are further elucidated by microscopy and spectroscopy techniques. Fig. 2(a) shows the optical image of a particular flake after an oxidization with $t_h$ = 1 h under the aforementioned conditions. Raman mapping with 488 nm laser excitation was done on the same flake and the light intensity was kept low to avoid any laser-assisted oxidation [14]. Fig. 2(b) plots the typical Raman shift data both at the edge and in the center of the flake. The prominent Raman peak at ~248 $cm^{-1}$ seen inside the flake can be attributed to the $E^1_{2g}$ mode of thin $WSe_2$ nano-sheets [20]. At the sample edges, several Raman peaks consistent with values reported for crystalline $WO_3$ [19, 21] were observed at around 134 $cm^{-1}$, 271 $cm^{-1}$, 713 $cm^{-1}$, and 805 $cm^{-1}$, respectively. The two images in the inset correspond to maps summing up the Raman intensity counts in the frequency intervals of 230 – 260 $cm^{-1}$ and 770 – 840 $cm^{-1}$, respectively. The Raman maps provide strong evidence that the optical contrast in Fig. 2(a) is indeed between $WO_3$ at the edges and $WSe_2$ in the center.

A microwave impedance microscope (MIM) based on commercial atomic-force microscopy (AFM) platforms [22] has been used to spatially resolve the local conductivity of the thermally oxidized WSe$_2$ nano-sheets. The measurement does not need contact electrodes and is generally non-invasive to the materials. As a low-frequency equivalent of near-field scanning optical microscopy (NSOM), the MIM has a spatial resolution defined by the tip radius (10~100 nm) rather than the wavelength of the microwave [23]. Here, a 1 GHz excitation signal is guided to the tip apex through the center conductor of a shielded cantilever probe [24]. During the scanning, the MIM electronics measure the real and imaginary parts of the tip-sample admittance (reciprocal of impedance) to form the corresponding MIM-Re and MIM-Im images. The MIM-Im signal increases monotonically as a function of the sample conductivity. The MIM-Re signal, on the other hand, peaks at intermediate conductivities and decreases for both good metals and good insulators. Detailed descriptions of the MIM system can be found in Ref. [25].

Figs. 2(d–f) show the simultaneously acquired AFM and MIM images of the same sample in Fig. 2(a). Compared with the as-exfoliated sample, the oxidation resulted in a moderate decrease of its thickness from 30 nm to 28 nm at the edges. On the other hand, the MIM images reveal high electrical conductivity in certain locations of the flake. First, large MIM-Im signals were observed at the WO$_3$–WSe$_2$ interface. Secondly, bright spots (high signals) randomly distributed in the interior of the flake were also seen in the MIM-Im image. The high local conductivity is presumably due to the partial oxidation of WSe$_2$ into sub-stoichiometric WO$_{3-x}$, which has been reported to be a good electrical conductor at the room temperature [26]. The MIM images, showing these partially oxidized regions, thus provide spatially resolved information of the oxidation process. By comparison with the optical and Raman data in Figs. 2(a) and 2(b), it is obvious that the high conductivity occurred within the regions where the bulk sample remained mostly WSe$_2$. Considering the secondary ion mass spectroscopy data to be detailed below, we further surmise that partial oxidation in the center area mainly took place at the top surface. Similar MIM responses were also measured in thinner WSe$_2$ nano-sheets down to 2–4 nm (Supporting Information S3). For comparison, we have also annealed WSe$_2$ flakes at lower temperatures, e.g., 200 ºC and 300 ºC. As seen in Supporting Information S4, the edge-initiated oxidation process was again observed, while the reaction rate became slow at 300 °C and negligible at 200 °C.

In order to further understand the spatial progression of oxidation, we repeatedly oxidized WSe$_2$ samples at 400 °C for a number of steps with the same heating/cooling profile described before.

Room-temperature MIM data were acquired in between these steps. Fig. 3(a) displays selected MIM-Im and MIM-Re images of a 45-nm thick flake during this process, and the complete set of data can be found in Supporting Information S5. After the first cycle with $t_h$ = 2 min, the MIM maps show that the perimeter of the WSe$_2$ nano-sheets became conductive. After a second thermal cycle with $t_h$ = 8 min, the conductive regions appeared both around the edges and in the interior of the sample. For the next run of oxidation with $t_h$ = 30 min, the high conductivity at the edges started to diminish, indicative of the formation of fully oxidized WO$_3$. After two more steps with extended annealing time over 2 hours, the entire sample became a good insulator, as indicated from the low MIM signals in both channels.

With the goal of quantitative understanding of the MIM data, we performed finite-element analysis (FEA) modeling [25] to estimate the local sheet resistance $R_{sh}$ at various regions of the thermally oxidized samples. Details of the simulation based on the actual sample geometry are provided in Supporting Information S6. The simulated MIM response is plotted in Fig. 3(b), from which $R_{sh}$ at different regions can be extracted. For the conductive WO$_{3-x}$, the MIM signals (~ 4 V in MIM-Im and ~ 1 V in MIM-Re) correspond to $R_{sh}$ ~ $10^5$ Ω/sq. A small MIM-Re signal was seen for the as-exfoliated semiconducting WSe$_2$, corresponding to $R_{sh}$ ~ $10^8$ Ω/sq. In contrast, the MIM-Re signal was negligible for the fully oxidized WO$_3$, indicative of very high sheet resistance beyond $10^9$ Ω/sq. Note that for the highly resistive WSe$_2$ and the insulating WO$_3$ regions, the dielectric constants of both materials [10 – 12, 27, 28] can also be estimated to be around 16 from the FEA modeling [29, 30]. We have also performed transport studies on micro-fabricated devices to measure the DC resistances (Supporting Information S7). The estimated sheet resistances were on the order of $10^5$, $10^8$, and $10^9$ Ω/sq for WO$_{3-x}$, WSe$_2$, and WO$_3$, respectively, in good agreement with the MIM results. The high local conductivity in the partially oxidized regions, as shown by both the MIM and transport data, could lead to other interesting phenomena such as the plasmonic resonance reported in TMDC materials [31].

We now turn to the compositional analysis of oxidized WSe$_2$ using time-of-flight secondary ion mass spectroscopy (ToF-SIMS), a sample-destructive method that provides 3D elemental information of materials. Fig. 4 shows the results from two WSe$_2$ nano-sheets on the same SiO$_2$/Si substrates, which were oxidized for $t_h$ = 1 h at 400 °C and imaged by MIM before the ToF-SIMS measurement. For the relatively large flake (lateral size > 50 μm), the 3D rendering of the MIM-Im image in Fig. 4(a) again reveals highly conductive regions at the edges and several regions in the interior, where strong $WO_2^-$ (216 atomic mass units, amu) signals were detected in the ToF-SIMS experiment for the first few seconds of ion sputtering, i.e., at the surface of the nano-sheets. For longer sputtering time, the composition map in Fig. 4(b) shows mostly $Se^-$ (80 amu) signals in the center, i.e., inside the flake and deep below the surface. In Figs. 4(d) and 4(e), similar features were also observed in relatively small samples (lateral sizes < 20 μm), where substantial fractions of $WSe_2$ were oxidized. The depth profiles of $WO_2^-$ and $Se^-$ signals in Figs. 4(c) and 4(f) suggest that, within the resolution of ToF-SIMS, the widths of the peripheral $WO_3$ sections are nearly constant throughout the height of the flakes. Finally, an intriguing observation in Fig. 4 is that the $WO_2^-$ signal was also detected on the entire bottom surface, which may be driven by the water molecules trapped in the $WSe_2$–$SiO_2$ interface at the time of mechanical exfoliation. In fact, with a high-humidity pretreatment of the samples, it can be shown that water molecules on the top surface can lead to extensive oxidation here (Supporting Information S8). At this stage, it is difficult to separately determine the structural and electrical properties of such a buried layer from the rest of the sample. Further experiments with controlled substrate pretreatments and different substrates are needed in order to understand the origin of such extensive oxidation at the bottom surface, as well as its contribution to the overall electrical properties during the oxidation process.

To summarize, the thermal oxidation of thin nano-sheets of $WSe_2$ progresses in several different ways. The dominant component of this process starts at the flake edges and moves towards the center, as indicated by optical and Raman images. The near-field microwave impedance microscopy, with better spatial resolution than optical techniques, shows that the intermediate reaction product at the $WO_3$–$WSe_2$ interface and certain regions at the surface can have a much higher local conductivity than either $WSe_2$ or $WO_3$. The oxidation also occurs at the bottom surface, which may be related to water trapped between the exfoliated $WSe_2$ and the substrate. Nanoscale mapping of chemical reactions could be of particular importance for 2D materials in terms of their potential for use in electronics and other applications.

## Methods

**Raman mapping.** Raman spectroscopy was done with a WiTec Alpha 300 micro-Raman confocal microscope, with a laser wavelength of 488 nm and a grating of 1800 lines/mm. An integration time of 1 sec was used with a mapping resolution of 4 pixels/micron.

**Transmission electron microscopy.** Cross sectional samples were prepared by first covering with an epoxy protection layer (EpoxyBond 110, Allied High Tech) and then cutting with a dicing saw (DISCO DAD-321). The samples were then protected with a Pt layer deposited with a focused ion beam system (FEI TEM 200), which was then used to thin the areas of interest. The samples were characterized with a JEOL 2010F transmission electron microscope.

**Device characterization.** Devices with ~ 40 nm Pd contacts were fabricated using electron beam lithography with a JEOL 6000 FSE system. Microchem A4 950 Poly (methyl methacrylate) (PMMA) applied at 4000 rpm for 40 sec was used as the resist. Contacts were made using electron beam evaporation, and the lift-off was done in acetone. The properties were measured under room temperature in a vacuum pressure of $< 10^{-4}$ Torr with a Lake Shore model CPX cryogenic probe station and an Agilent 4156C semiconductor parameter analyzer. The device data can be found in the Supporting Information S7.

**Time-of-flight secondary ion mass spectroscopy.** The characterization was done using a ToF-SIMS (ION-ToF GmbH ToF-SIMS 5) configured with 1000 eV $Cs^+$ sputtering and 30 keV $Bi^+$ analysis ion beams. The ToF-SIMS primary ion gun was configured to be in the acquisition and burst alignment (BA) mode ($Bi_1^+$ at 30 keV ion energy with a pulse duration of 100 ns without bursts) for high lateral resolution.

**Microwave impedance microscopy.** The MIM in this work is based on a standard AFM platform (ParkAFM XE-70). The customized shielded cantilevers are commercially available from PrimeNano Inc. Finite-element analysis was performed using the commercial software COMSOL4.4. Details of the MIM experiments and numerical modeling can be found in the Supporting Information S6.

## Acknowledgements

We thank Llewellyn K. Rabenberg, Federico Cardenas, Hugo Celio, and Wei Jiang for valuable discussions. We thank Jung-Fu Lin and Sanjay Banerjee for allowing us the use of their experimental setups. We also thank Karalee Jarvis for assistance with TEM and Maria Hall for TEM sample preparation. The MIM work (Y.L., D.W., K.L.) was supported by Welch Foundation Grant F-1814. C. T. acknowledges support from a National Defense Science and Engineering Graduate (NDSEG) Fellowship: Contract FA9550-11-C-0028, awarded by the Department of Defense. Work by C.T., A.N., H.Y.C., J.K., and D.A. is supported in part by the Army Research Office (ARO), the Office of Navy Research (ONR), the Southwest Academy of Nanoelectronics (SWAN), a Nanoelectronics Research Initiative (NRI) center. H.C., Y.H., R.P., and R.S.R. were sponsored in part by ONR. RSR was also supported by IBS-R019-D1. X.W. acknowledges support from the Nanomanufacturing Systems for Mobile Computing and Mobile Energy Technologies (NASCENT). R.G. acknowledges support from Center for Low Energy Systems Technology (LEAST), one of six centers supported by the STARnet phase of the Focus Center Research Program (FCRP), a Semiconductor Research Corporation program sponsored by MARCO and DARPA.

## Additional information

The authors declare no competing financial interests. Supplementary information is available online.

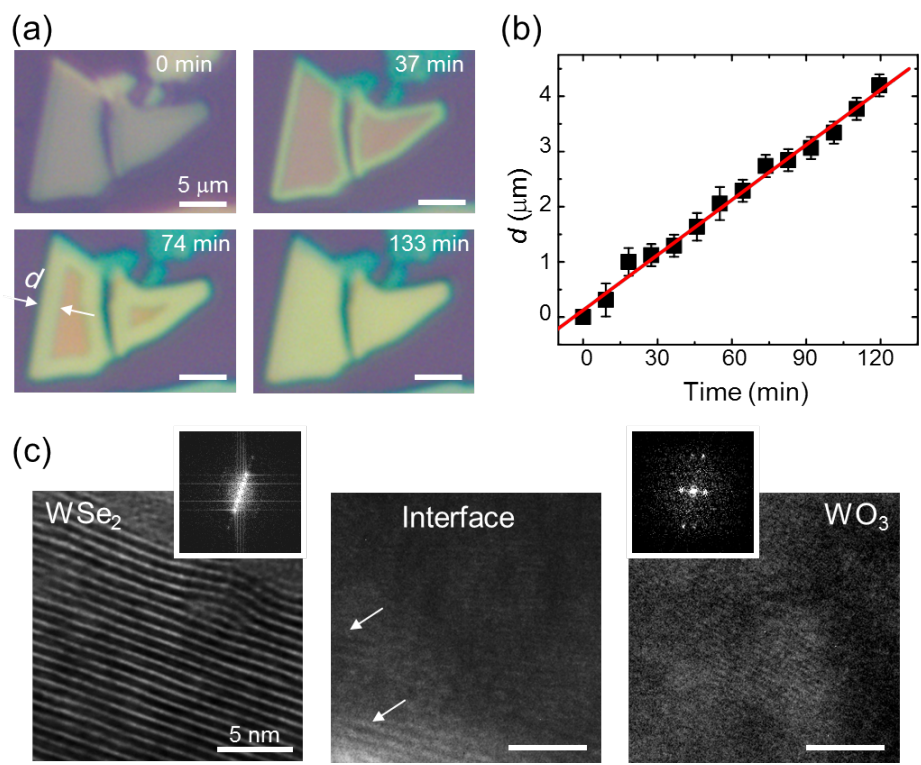

**Figure 1.** (a) Representative pictures captured by a 20× microscope objective during a continuous thermal oxidation process. All scale bars are 5 μm. (b) Lateral distance from the sample edge to the optically measured $WO_3$–$WSe_2$ interface as a function of time. A linear dependence is observed. (c) Cross sectional TEM images of the $WSe_2$ (left), $WO_3$ (right), and the partially oxidized interface (middle). Remnant features of the van der Waals structure, as indicated by arrows, can still be seen at the $WO_3$–$WSe_2$ interface. The insets are FFT images of the raw data. All scale bars in (c) are 5 nm.

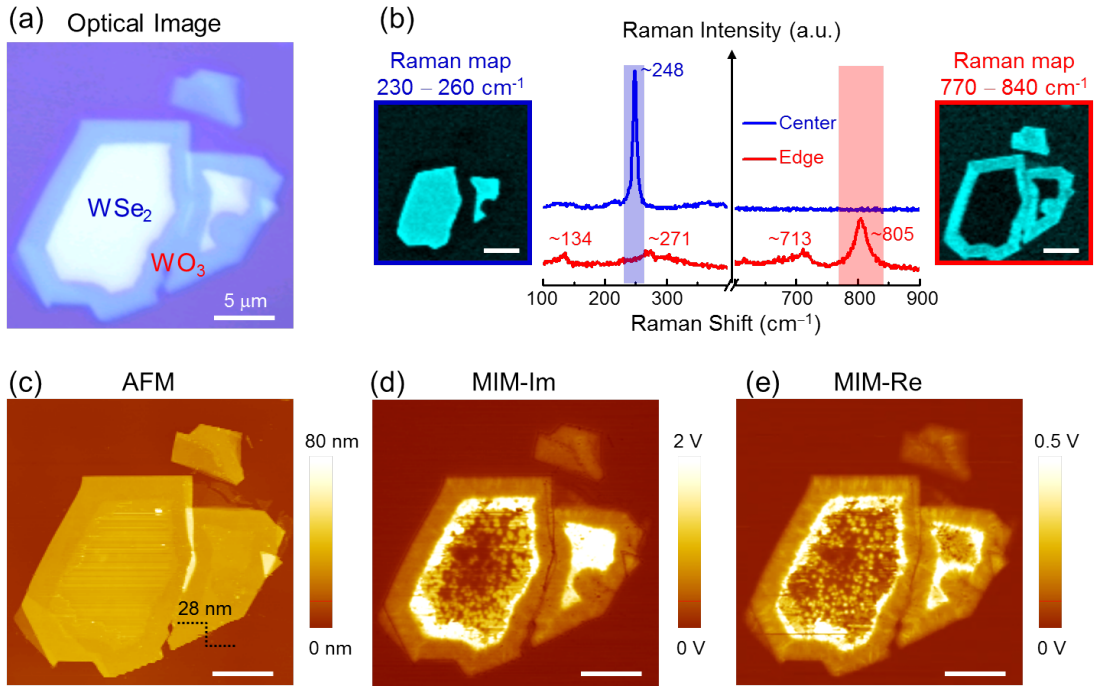

**Figure 2.** (a) Optical microscopy (OM) picture of WSe$_2$ nano-sheets taken under a 100× microscope objective. (b) Typical Raman intensity plots taken at the edge (red) and center (blue) of the flakes. The two Raman maps in the inset correspond to the sum of Raman intensity counts in the range of 230 – 260 cm$^{-1}$ (blue) and 770 – 840 cm$^{-1}$ (red). (c) AFM, (d) MIM-Im, and (e) MIM-Re images of the same sample described above. The highly conductive regions are brighter in the MIM maps. All scale bars are 5 μm.

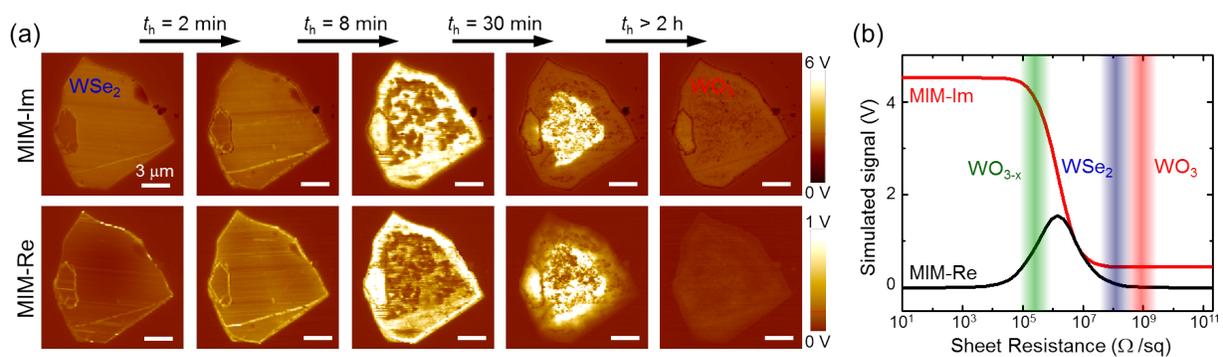

**Figure 3.** (a) MIM images acquired in between several heating/cooling steps. The lengths of time $t_h$ for which the sample was held at 400 °C are shown above the images. All scale bars are 3 μm. (b) Simulated MIM-Im (red) and MIM-Re (black) signals as a function of the sheet resistance of a 45 nm-thick flake. The range of measured MIM signals on $WO_{3-x}$ (green), $WSe_2$ (blue), and $WO_3$ (red) regions are indicated in the plot.

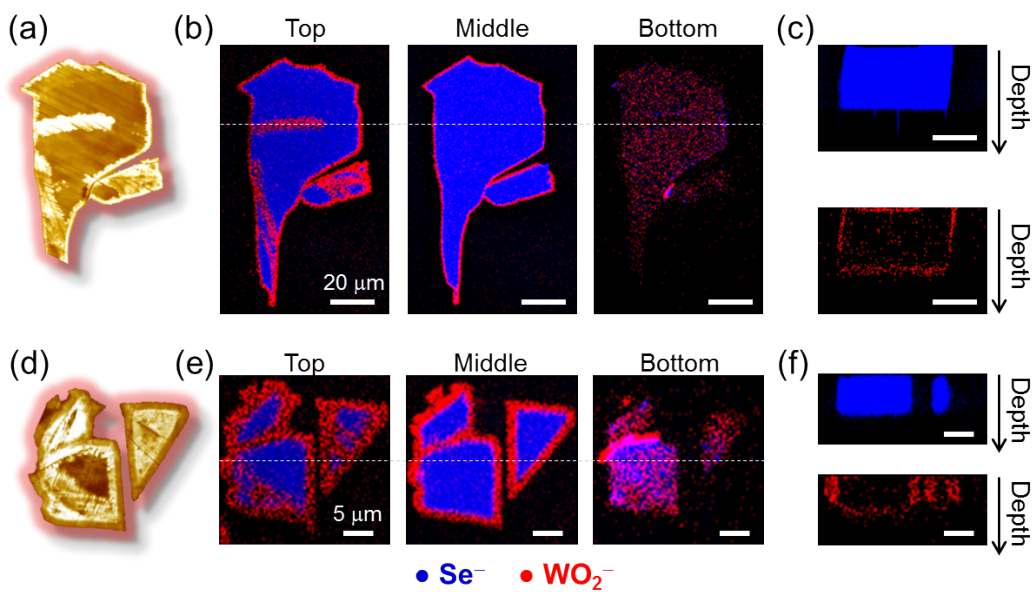

**Figure 4.** (a) 3D rendering of the MIM-Im image of a large $WSe_2$ sample (thickness ~ 140 nm) after one hour of thermal oxidation. (b) Time-of-flight secondary ion mass spectroscopy (ToF-SIMS) maps at the beginning (left), middle (center), and end (right) of the experiment, during which the nano-sheet was completely etched away by the ion sputtering. The false colors of blue and red corresponds to $Se^-$ (80 amu) and $WO_2^-$ (216 amu), respectively. (c) ToF-SIMS depth profile in two orthogonal directions of the same region. (d – f) MIM and ToF-SIMS data, as (a – c), of several smaller sheets (thickness ~ 48 nm). The scale bars are 20 μm in (b, c) and 5 μm in (e, f).

ToC image:

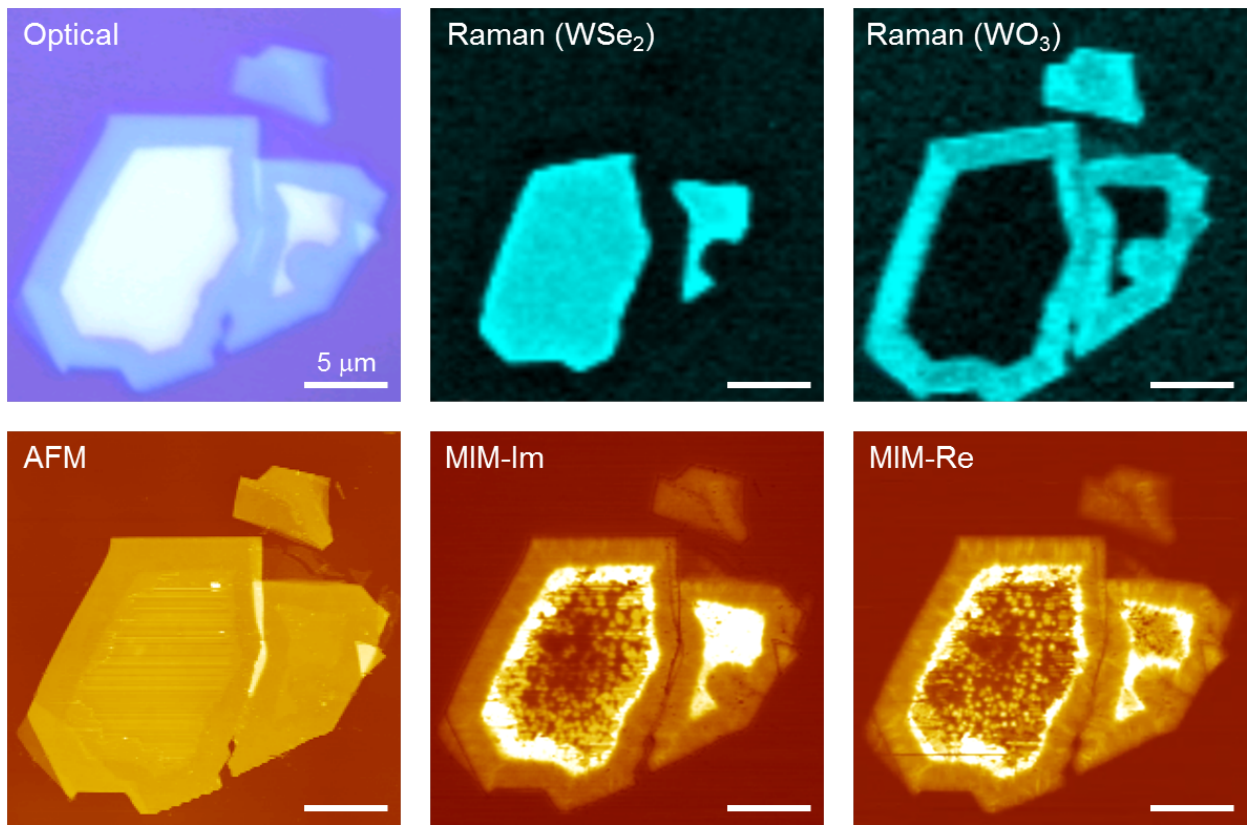

# Supporting Information

## Thermal Oxidation of WSe$_2$ Nano-sheets Adhered on SiO$_2$/Si Substrates

Yingnan Liu,[a,‡] Cheng Tan,[b,c,‡] Harry Chou,[c] Avinash Nayak,[b] Di Wu,[a] Rudresh Ghosh,[c] Hsiao-Yu Chang,[b] Yufeng Hao,[c] Xiaohan Wang,[c] Joon-Seok Kim,[b] Richard Piner,[c] Rodney S. Ruoff,[c,d,e] Deji Akinwande,[b,]* Keji Lai[a,]*

a) Department of Physics, University of Texas at Austin, Austin, TX, 78712, USA
b) Microelectronics Research Center, University of Texas at Austin, Austin, TX 78758, USA
c) Department of Mechanical Engineering and the Materials Science and Engineering Program, University of Texas at Austin, Austin, TX 78712, USA
d) Center for Multidimensional Carbon Materials, Institute for Basic Science (IBS), Ulsan 689-798, Republic of Korea
e) Department of Chemistry, Ulsan National Institute of Science and Technology (UNIST), Ulsan 689-798, Republic of Korea

* Correspondence and requests for materials should be addressed to K.L. (kejilai@physics.utexas.edu) or D.A. (deji@ece.utexas.edu)

‡ These authors contributed equally to this work.

**Section 1: Temperature profile of the thermal oxidation process.**

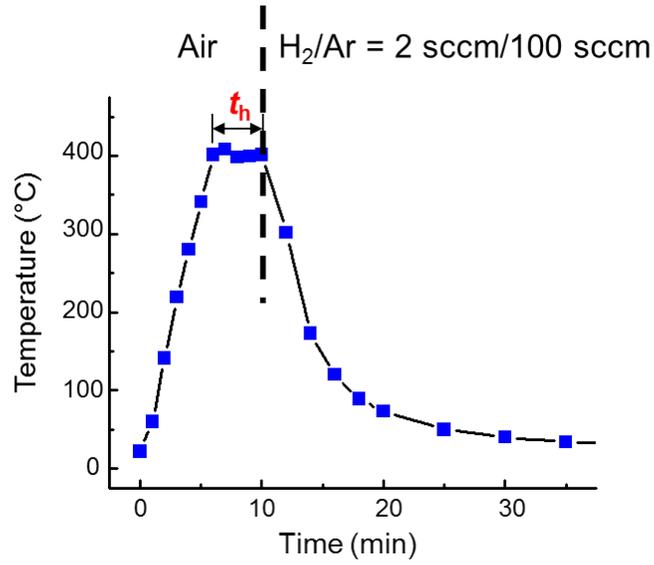

Fig. S1. Temperature-time profile of the heating/cooling process in this work.

Fig. S1 shows a typical temperature-time profile of the thermal oxidation of WSe$_2$ nano-sheets in this work. The samples were heated up in a quartz tube furnace to $T = 400$ ºC with a ramp time of 6 minutes, and then annealed in air at 400 ºC for a holding time of $t_h$, in this case 4 min. The heater was then turned off and the furnace was purged with H$_2$/Ar (2 sccm/100 sccm). The sample temperature rapidly dropped below 100 ºC in less than 10 min.

**Section 2: Thermal oxidation of WSe$_2$ nano-sheets as a function of the heating time.**

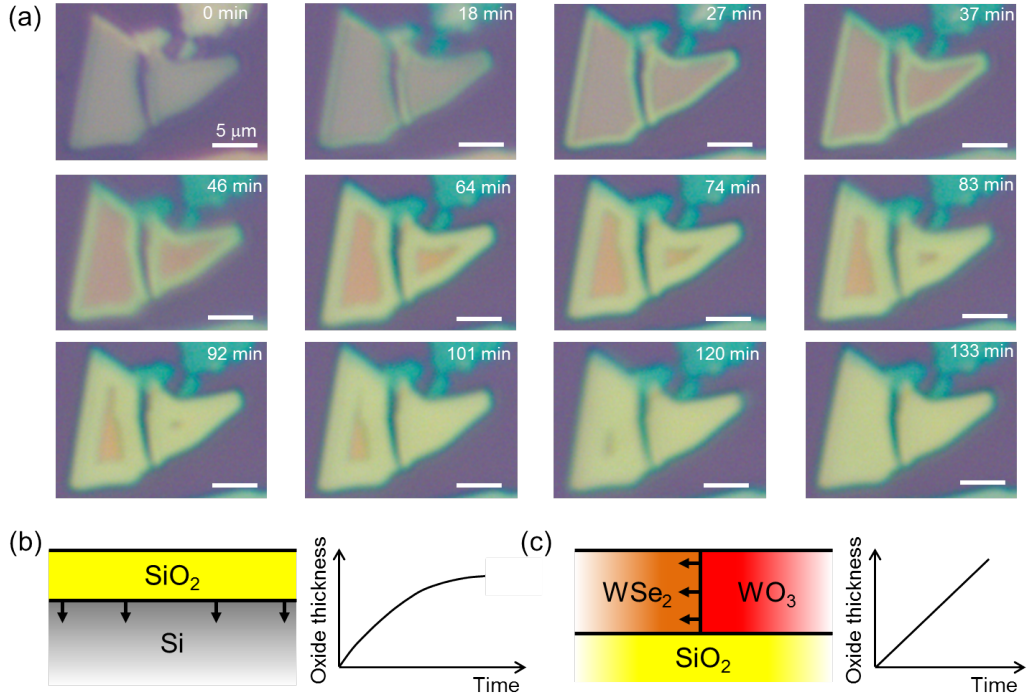

Fig. S2. (a) A series of optical images captured by a video camera during a continuous thermal oxidation of WSe$_2$ nano-sheets. All scale bars are 5 μm. (b) Schematic of the vertical propagation of thermal oxides in Si. The corresponding thickness-time dependence shows an initial linear regime and a transition to the quadratic regime for long enough time. (c) Schematic of the horizontal propagation of thermal oxides in WSe$_2$, corresponding to a linear thickness-time dependence.

Optical images recorded by a video camera with a 20× microscope objective are shown in Fig. S2(a) during a continuous thermal oxidation of WSe$_2$ nano-sheets. In conventional 3D semiconductors such as Si, the oxidation requires the diffusion of oxidants (O$_2$ or H$_2$O) through the existing oxides to the oxide-substrate interface before reaction with the substrate. According to the Deal-Grove model [S1], this process involves an initial linear regime at which the interface reaction dominates, and a transition to the quadratic regime for long enough time, at which the diffusion of oxidants through the thickening oxide layer dominates [Fig. S2(b)]. For the oxidation of WSe$_2$, however, oxidants can be supplied directly to the interface. As depicted in Fig. S2(c), the width of the oxide displays the linear time dependence observed in Fig. 1(b), indicating that the oxidation is limited by the chemical reaction at the interface rather than diffusion through the oxides. Note that the nearly vertical WSe$_2$–WO$_3$ interface is observed in the ToF-SIMS data shown in Fig. 4.

**Section 3: Thermal oxidation of few-layer WSe$_2$ samples.**

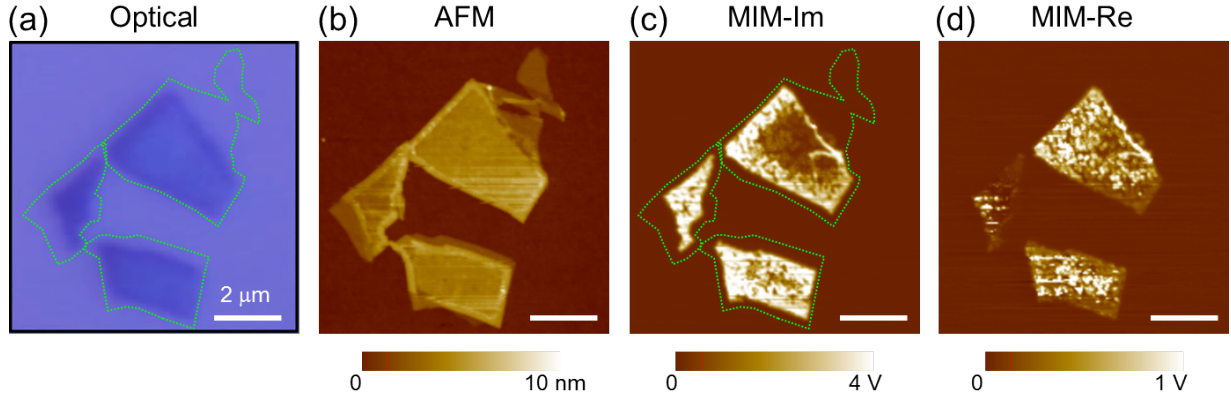

Fig. S3. (a) Optical, (b) AFM, (c) MIM-Im, and (d) MIM-Re images of thin WSe$_2$ samples after the thermal oxidation. All scale bars are 2 μm.

Optical, AFM, and MIM images of few-layer WSe$_2$ (2 ~ 4 nm in thickness) samples after a short ($t_h$ = 1 min) thermal oxidation are shown in Fig. S3. In general, the results are qualitatively the same as that observed in thick samples (Fig. 2). The thinnest (2 nm) regions and part of the edges are invisible in the optical and MIM images, presumably due to the insulating behavior of fully oxidized WO$_3$. A drastic increase of conductivity was observed at the perimeter of the flakes, extending ~ 1 μm from the WO$_3$–WSe$_2$ interface towards the center. Large MIM signals, indicative of higher local conductivity, can also be seen in the interior of the nano-sheets.

**Section 4: Thermal oxidation of WSe$_2$ samples at lower temperatures.**

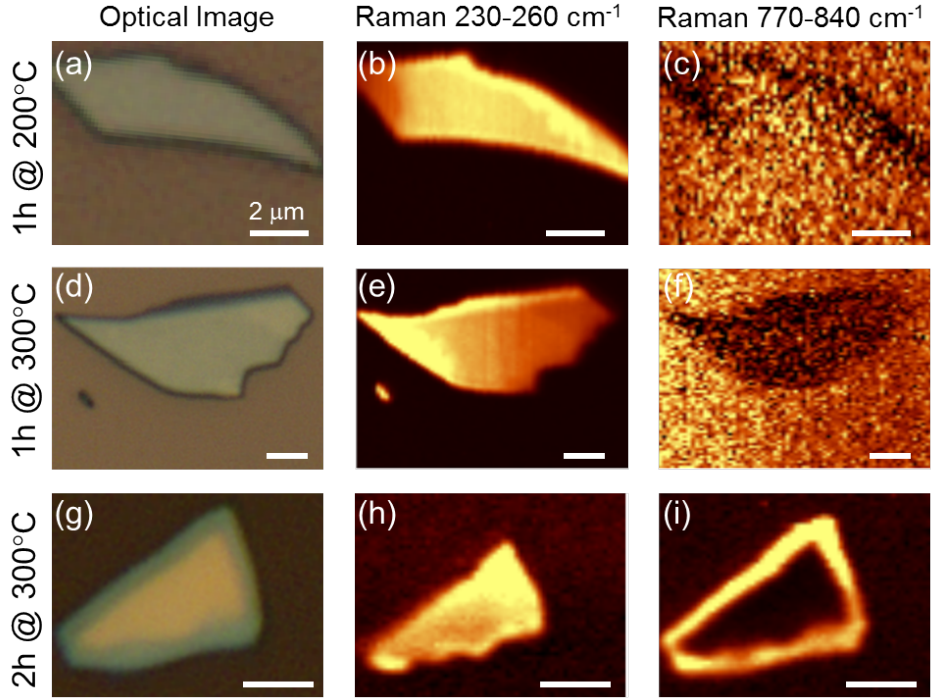

Fig. S4. (a – c) Optical and Raman maps in the range of 230 – 260 cm$^{-1}$ (WSe$_2$) and 770 – 840 cm$^{-1}$ (WO$_3$) of a WSe$_2$ nano-sheet after 1 hour thermal oxidation at 200 ºC. (d – f) and (g – i) Similar results for 1 hour and 2 hours thermal oxidation at 300 ºC, respectively. All scale bars are 2 μm.

For comparison with the thermal oxidation at 400 °C, we have also carried out the annealing experiment at lower temperatures, e.g., 200 °C and 300 °C, which are commonly used to remove tape residues during the exfoliation before making electronic devices. Note that for that purpose, the annealing is performed inside vacuum or with the flow of inert gas, in contrast to the oxygen environment in our oxidation experiment. As shown in Fig. S4 (a – f), no obvious oxidation was observed from Raman and optical images for samples annealed at 200 ºC and 300 ºC for an hour. For longer oxidation time of 2 hours at 300 ºC, the edge-initiated oxidation was indeed observed, as shown in Fig. S4 (g – i). However, the rate of lateral oxidation at 300 ºC appears to one order of magnitude slower (~ 0.3 μm per hour) than that at 400 °C (~ 2 μm per hour). We therefore conclude that even with small oxygen contamination in the normal vacuum or inert gas environment, a brief annealing at 200 °C or 300 °C should not lead to severe oxidation of the WSe$_2$ samples.

## Section 5: Full set of AFM/MIM data for Figure 3.

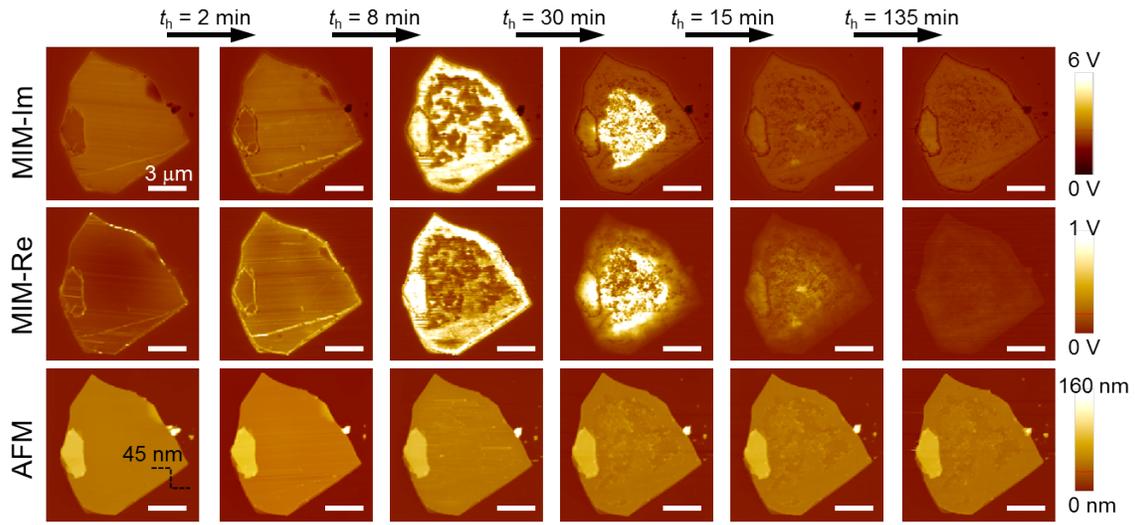

Fig. S5. Full set of AFM and MIM images shown in Fig. 3(a) in the main text. All scale bars are 3 μm.

The complete set of MIM and AFM images in Fig. 3(a) after each oxidation cycle is shown in Fig. S5. The holding time $t_h$ at 400 °C of each run is labeled on top of the images. As seen in the AFM data, the average sample thickness hardly changed throughout the entire process, although the oxidation indeed resulted in appreciable surface roughness.

**Section 6: Finite-element analysis of the thermally oxidized sample.**

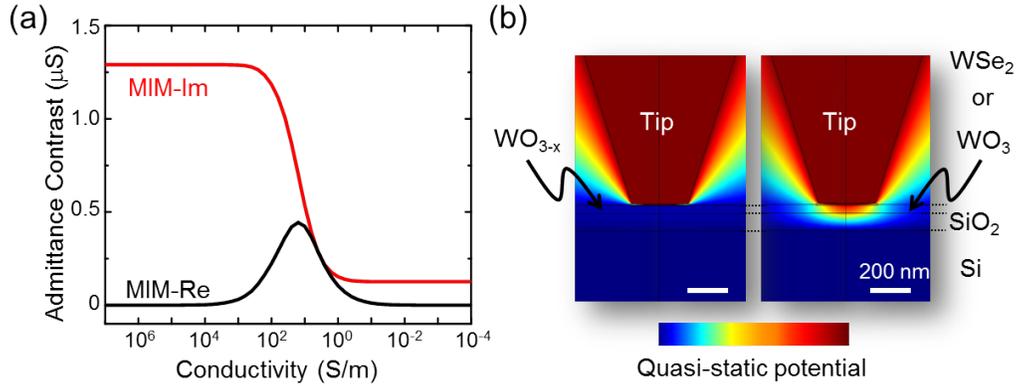

Fig. S6. (a) Finite-element modeling result on the thermally oxidized WSe$_2$ sample. (b) Quasi-static potential distribution when the tip is in contact with a thin flake of conductive WO$_{3-x}$ (left) or insulating WSe$_2$/WO$_3$ (right).

Fig. S6(a) shows the COMSOL4.4 FEA simulation result [S2] using the actual tip-sample dimensions and material properties. A blunt tip (diameter ~ 300 nm) due to repeated scans is assumed here, as indicated by a large signal on our standard calibration sample (not shown). For the highly resistive WSe$_2$ and the insulating WO$_3$ regions, the dielectric constants of both materials can also be estimated to be around 16 from the FEA modeling [S3, S4]. The values are consistent with those reported in the literature (WSe$_2$: $\varepsilon_\parallel$ = 4.2 or $\varepsilon_\perp$ = 12.8 or 18.5 [S5], WO$_3$: $\varepsilon$ = 12 or 14 for sintered samples and 20 for evaporated films [S6]). Fig. S6(b) shows the quasi-static potential around the tip when the sample is conductive (WO$_{3-x}$, left) or insulating (WSe$_2$ or WO$_3$). The FEA software computes the admittance contrast over the SiO$_2$/Si substrate as a function of the local conductivity $\sigma$. To compare with the transport data, we have converted the conductivity to sheet resistance (reciprocal of sheet conductance) $R_{sh}$ = $1/\sigma d$ in the main text, where $d$ = 45 nm is the sample thickness. Since both the lateral and vertical resolutions of the MIM are on the same order of the tip diameter, the technique cannot resolve the vertical conductivity distribution within the nano-sheet sample. The total gain of the MIM circuits is such that an admittance contrast of 1 µS corresponds to an output voltage of ~3.5 V. Note that several MIM probes with different tip diameters were used in this work, resulting in different signal strengths. The ratio between the MIM-Re and MIM-Im signals, on the other hand, is relatively insensitive to the tip condition and can be used to estimate the sheet resistances.

**Section 7: Transport studies on two-terminal devices.**

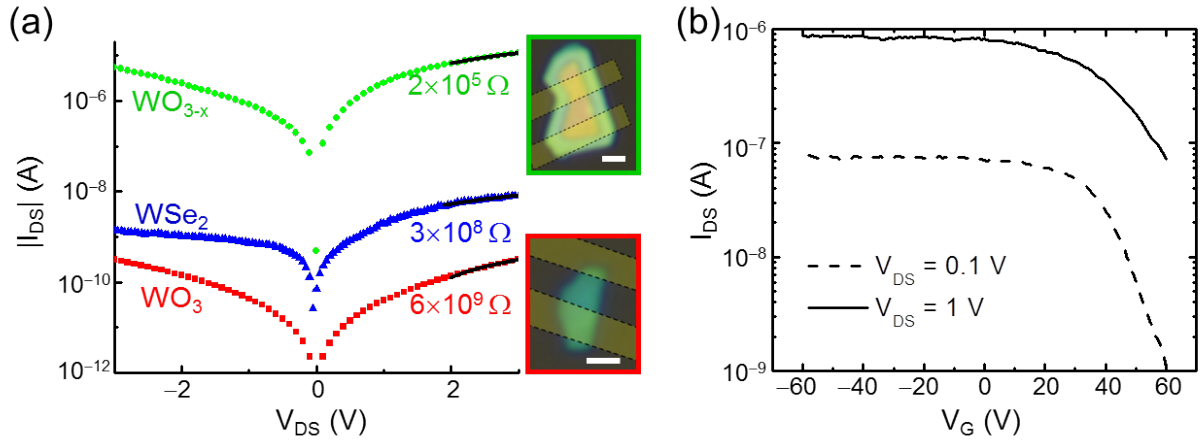

Fig. S7. (a) I-V characteristics of typical two-terminal devices. The blue, green, and red curves represent data from pristine $WSe_2$, partially oxidized $WO_{3-x}$, and fully oxidized $WO_3$ samples, respectively. The optical pictures of $WO_{3-x}$ and $WO_3$ samples are shown in the insets. Both scale bars are 1 μm. (b) Gate effect of a partially oxidized sample, showing clear p-type characteristics.

We have performed transport studies on micro-fabricated devices to measure the DC resistance. Fig. S7(a) shows the two-terminal I-V curves of three typical devices patterned on as-exfoliated $WSe_2$, partially oxidized $WO_{3-x}$, and completely oxidized $WO_3$ regions. While the $WSe_2$ and $WO_3$ samples were spatially uniform, electrodes on this $WO_{3-x}$ device inevitably covered a segment with strong electrical inhomogeneity. Fortunately, the result is still useful since the transport current will mostly go through the highly conductive areas. As plotted in Fig. S7(a), the rectified I-V characteristics at low bias voltages were due to the Schottky effect. For high enough bias voltages, the I-V curves became linear and the extracted two-terminal resistances are on the order of $10^5$ Ω, $10^8$ Ω, and $10^9$ Ω for $WO_{3-x}$, $WSe_2$, and $WO_3$, respectively. All devices have roughly one square in between the electrodes. Ignoring contact resistances at high biases, the electrical transport results, which provide an order-of-magnitude estimation of the sheet resistances, are in good agreement with the MIM data.

The gate dependence of a similar partially oxidized sample was also studied and the results are shown in Fig. S7(b). We caution that, due to the strong electrical inhomogeneity in partially oxidized $WSe_2$, it is difficult to assign the apparent p-type characteristics to the $WO_{3-x}$ or $WSe_2$ segments of the sample. More experiments are needed to clarify the carrier type in the conductive regions.

**Section 8: Effect of water pretreatment.**

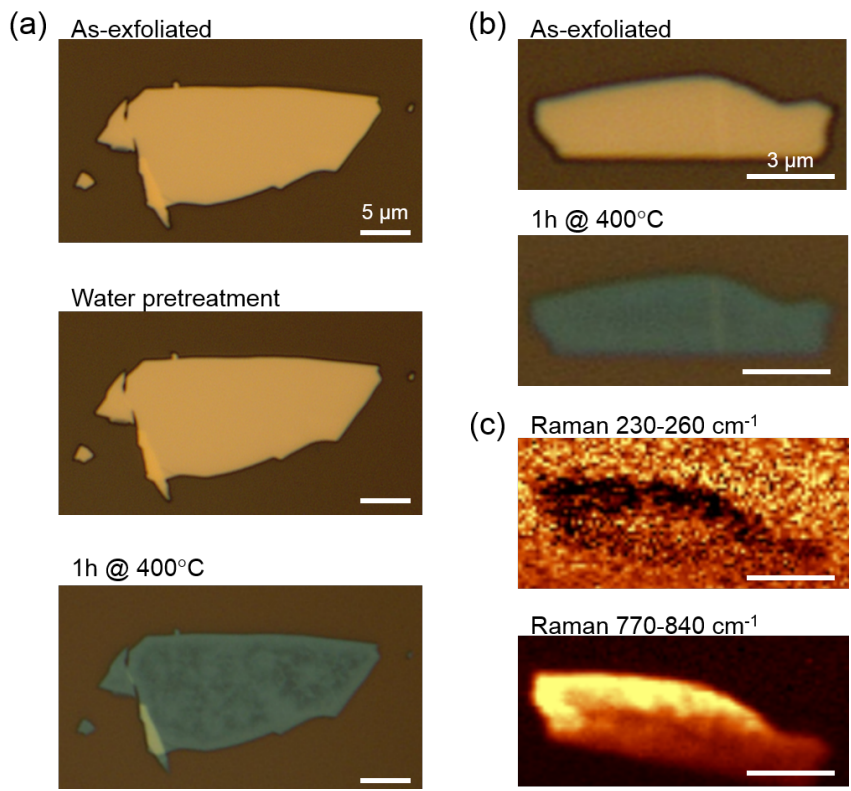

Fig. S8. (a) Optical images of a WSe$_2$ flake (top) right after the exfoliation, (middle) after water pretreatment, and (bottom) after oxidation at 400 °C for one hour. (b) Optical images of another WSe$_2$ flake after a similar process as (a). (c) Raman maps of the same flake as (b) in the range of 230 – 260 cm$^{-1}$ (WSe$_2$) and 770 – 840 cm$^{-1}$ (WO$_3$). The scale bars are 5 μm in (a) and 3 μm in (b) and (c).

In order to investigate the effect of trapped water molecules on the thermal oxidation, we have prepared samples in a highly humid (50 ~ 60% humidity) and warm (55 ~ 85 °C) environment for 90 minutes before the same annealing process described in S1. While the water pretreatment does not result in obvious changes before the annealing, the thermal oxidation process can be quite different from the edge-initiated one described in the main text. In fact, as indicated by both optical and Raman maps in Fig. S8, the entire top surface of the flakes after exposure to high humidity appears to be heavily oxidized, consistent with our speculation that water molecules trapped at the bottom surface may be responsible for the extensive oxidation. It is likely that the wet oxidation of WSe$_2$ resembles the process in the ozone environment [S7], where the reaction is dominated by the surfaces rather than the edges. Further experiments with controlled temperature, humidity level, and pretreatment time are needed to elucidate this point.